\documentclass[lettersize,journal]{IEEEtran}
\usepackage{amsmath,amsfonts}
\usepackage{algorithmic}
\usepackage{algorithm}
\usepackage{array}
\usepackage[caption=false,font=normalsize,labelfont=sf,textfont=sf]{subfig}
\usepackage{textcomp}
\usepackage{stfloats}
\usepackage{url}
\usepackage{verbatim}
\usepackage{graphicx}
\usepackage{cite}
\usepackage{multirow}
\usepackage{makecell}
\usepackage{bbding}
\usepackage{utfsym}
\usepackage{balance}
\DeclareUnicodeCharacter{2212}{\ensuremath{-}}

\usepackage{amssymb}
\usepackage{mathtools}
\usepackage{amsthm}

\usepackage{hyperref}
\usepackage{cite}
\usepackage{booktabs}
\def\rvx{{\mathbf{x}}}

\def\rvv{{\mathbf{v}}}
\def\rvu{{\mathbf{u}}}

\begin{document}

\title{SimpleSpeech 2: Towards Simple and Efficient Text-to-Speech with Flow-based Scalar Latent Transformer Diffusion Models}
\author{Dongchao Yang, Rongjie Huang, Yuanyuan Wang, Haohan Guo, Dading Chong, Songxiang Liu, \\
Xixin Wu, Helen Meng 
\thanks{Dongchao Yang, Rongjie Huang, Yuanyuan Wang, Haohan Guo,   Xixin Wu, and Helen Meng are with the Chinese University of Hong Kong.} 
\thanks{Songxiang Liu and Dading Chong are independent researcher.}
\thanks{Helen Meng is the corresponding author.} }
%
\markboth{Journal of \LaTeX\ Class Files,~Vol.~14, No.~8, August~2021}%
{Shell \MakeLowercase{\textit{et al.}}: A Sample Article Using IEEEtran.cls for IEEE Journals}


\maketitle
\begin{abstract} 
Scaling Text-to-speech (TTS) to large-scale datasets has been demonstrated as an effective method for improving the diversity and naturalness of synthesized speech. At the high level, previous large-scale TTS models can be categorized into either Auto-regressive (AR) based (\textit{e.g.}, VALL-E) or Non-auto-regressive (NAR) based models (\textit{e.g.}, NaturalSpeech 2/3). Although these works demonstrate good performance, they still have potential weaknesses. For instance, AR-based models are plagued by unstable generation quality and slow generation speed; meanwhile, some NAR-based models need phoneme-level duration alignment information, thereby increasing the complexity of data pre-processing, model design, and loss design. In this work, we build upon our previous publication by implementing a simple and efficient non-autoregressive (NAR) TTS framework, termed SimpleSpeech 2. SimpleSpeech 2 effectively combines the strengths of both autoregressive (AR) and non-autoregressive (NAR) methods, offering the following key advantages: (1) simplified data preparation; (2) straightforward model and loss design; and (3) stable, high-quality generation performance with fast inference speed.
Compared to our previous publication, we present ({\romannumeral1}) a detailed analysis of the influence of speech tokenizer and noisy label for TTS performance; ({\romannumeral2}) four distinct types of sentence duration predictors; ({\romannumeral3}) a novel flow-based scalar latent transformer diffusion model. With these improvement, we show a significant improvement in generation performance and generation speed compared to our previous work and other state-of-the-art (SOTA) large-scale TTS models. Furthermore, we show that SimpleSpeech 2 can be seamlessly extended to multilingual TTS by training it on multilingual speech datasets. Demos are available on: {https://dongchaoyang.top/SimpleSpeech2\_demo/}.

\end{abstract}

\begin{IEEEkeywords}
Text-to-speech, Audio Codec, Audio generation, diffusion model
\end{IEEEkeywords}

\section{Introduction}
Text-to-Speech synthesis (TTS) endeavors to synthesize speech that is both intelligible and natural at a human level, a field that has witnessed significant advancements over the past few years due to the proliferation of deep learning technologies \cite{tecatron2,fastspeech2,vits,glow-tts,grad-tt,transformer-tts,ns1}. These TTS models have successfully synthesized speech that is not only intelligible but also of high quality, albeit generally limited to specific styles, speakers, and languages. This limitation primarily stems from the fact that they are trained on small-scale, high-quality, labeled speech datasets.
At a high level of modeling, previous TTS models are usually categorized into two types: auto-regressive (AR) and non-auto-regressive (NAR) models. Generally, NAR-based models outperform AR-based models in terms of generation speed and robustness \cite{fastspeech2}. However, AR-based models exhibit superior diversity, prosody, expressiveness, and flexibility, attributable to their implicit duration modeling and AR sampling strategy \cite{tecatron2}. Furthermore, many NAR-based models \cite{fastspeech2} depend on fine-grained alignment information (phoneme-level duration), which complicates data pre-processing. Although numerous studies \cite{grad-tt,glow-tts,vits} have suggested employing the Monotonic Alignment Search (MAS) strategy to learn alignment information, this approach not only increases training complexity but also results in alignments that may not always be suitable. Most critically, the rigid boundary between phoneme and speech representation can lead to unnatural prosody \cite{seedtss}.

Recently, an increasing number of researchers have concentrated on synthesizing diverse and natural speech, particularly focusing on the diversity of speaking styles, achieving human-level naturalness, and zero-shot speaker cloning. To achieve these objectives, they propose utilizing large-scale datasets to train Text-to-Speech (TTS) models \cite{borsos2023audiolm,valle,speartts,uniaudio,make-a-voice}.
For example, language model (LM)-based TTS systems, such as VALL-E \cite{valle}, employ a pre-trained audio codec model (Encodec \cite{encodec}) to convert the speech signal into a sequence of discrete tokens. Subsequently, an auto-regressive (AR) language model is trained to predict these speech tokens based on phonemes and speaker prompts. LM-based TTS models streamline data pre-processing, and their structure and training strategies draw from the successes of large language models in natural language processing (NLP) \cite{llama2,gpt4}. Moreover, these models are capable of generating expressive and diverse speech, thanks to their implicit duration modeling and sampling strategy. However, they suffer from slow and unstable generation due to the nature of AR sampling. To mitigate these drawbacks, large-scale non-autoregressive (NAR) models \cite{soundstorm,ns2,voicebox,megatts} have been proposed, such as NaturalSpeech 2/3. Although NAR-based large-scale TTS models exhibit higher inference speeds and enhanced stability, they require extra effort in data pre-processing, such as using their internal alignment tools to secure the alignment between speech and phonemes \cite{ns2,ns3}. Additionally, the persistent issue of hard boundary alignment continues to limit the natural prosody in NAR-based models \cite{seedtss}. We summarize the strengths and potential weaknesses for current AR-based and NAR-based large-scale TTS models in Table \ref{tab1}. We can see that both AR-based and NAR-based large TTS models still have considerable room for improvement in terms of performance, efficiency, or data requirements. Consequently, the primary research question is \textbf{whether we can develop a simple, efficient, and stable TTS system capable of synthesizing diverse, natural, and controllable speech}.
\begin{figure*}[t]
    \centering
    \includegraphics[width=\textwidth]{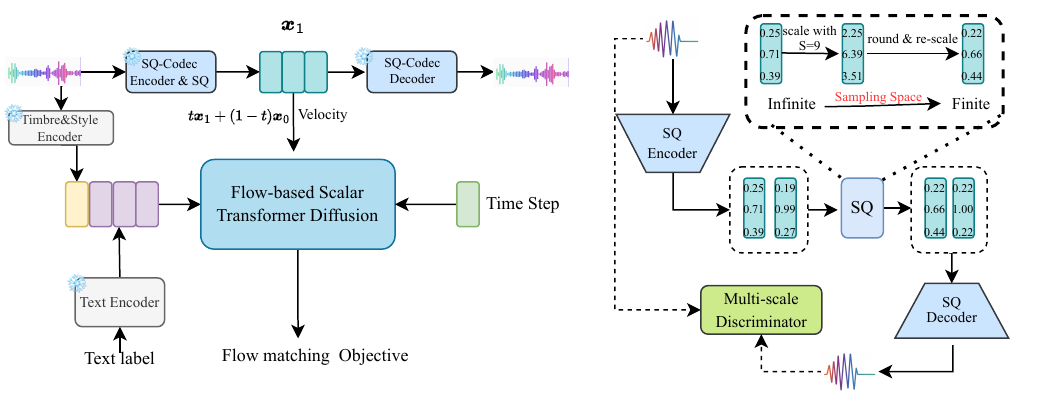}
    \vspace{-5mm}
    \caption{The left picture presents an overview of SimpleSpeech 2, while the right picture illustrates the SQ-Codec. It is important to note that the modules marked with a snowflake are frozen during the training of the flow-based scalar transformer diffusion. Detailed descriptions of the Timbre\&Style encoder, SQ-Codec, text encoder, and flow-based transformer diffusion are provided in Section \ref{sec4:Proposed method}.} 
    \label{fig:overview}
\end{figure*}

Considering the complexity and long sequence of the speech signal, most large TTS models use a speech tokenizer \footnote{In this study, we refer to the intermediate representation extractor as a speech tokenizer. In the audio field, the speech tokenizer can be a codec model or Variational Auto Encoder (VAE).} to extract the intermediate representation from speech, \textit{e.g.} discrete tokens of audio codec model \cite{valle,uniaudio} or continuous embeddings from VAE models \cite{ns2}, then applying generative models on the intermediate representation. However, previous models pay more attention to improving the performance of TTS systems by scaling the training data and model size. The research about the influence of speech tokenizers on the TTS models is scarce. In other words, previous works tell us how to utilize a pre-trained speech tokenizer for large TTS systems, but do not disclose what is the most suitable speech tokenizer for TTS systems. We think the speech tokenizer is a key point for the success of large TTS models. Thus, the second research problem is \textbf{what is a good speech tokenizer for large TTS models}. In this study, we propose to focus on two aspects of the speech tokenizer: 
\begin{itemize}
\item 
Completeness: The speech tokenizer should recover the compressed speech without loss.
\item 
Compactness: The speech tokenizer should encode the speech into fewer parameters or a smaller space.
\end{itemize}
In Section \ref{sub:good tokenizer}, we first give a comprehensive performance comparison for different speech tokenizers, including our proposed SQ-Codec, Encodec, and Variational Auto Encoder (VAE). Then we give a detailed analysis of the influence of completeness and compactness on speech generation. 

Large-scale datasets are crucial for the development of advanced Text-to-Speech (TTS) systems \cite{valle,ns2,ns3,uniaudio,voicebox,voicecraft}. While prior studies have demonstrated that large-scale Automatic Speech Recognition (ASR) datasets can be employed to train TTS systems, the details of obtaining text-speech pairs are usually neglected. In this study, we utilize publicly available ASR models (Whisper \cite{whisper}), to generate the transcriptions. However, these transcriptions may be inaccurate and are thus considered noisy labels. The third research question we address is: \textbf{Can a dataset comprising noisy labels effectively train a high-quality TTS system?} In Section 
\ref{sec:cfc}, we give a theoretical analysis that when a large-scale dataset includes a small number of noisy labels, which is equal to introducing the classifier-free guidance training \cite{cfc} for the model optimization.

Our contributions can be summarized as follows:
\begin{itemize}
\item  We introduce SimpleSpeech 2, a simple and efficient NAR framework for text-to-speech, building upon our previous work \cite{simplespeech}. This framework effectively combines the advantages of current AR-based and NAR-based models. 

\item  We conduct a comprehensive series of contrast experiments to explore the impact of different tokenizers on TTS systems, and investigate the influence of completeness and compactness of speech tokenizer for TTS performance.

\item We explore the influence of sentence duration predictor, and show 4 distinct duration predictors.

\item We propose a novel flow-based scalar transformer diffusion model that is faster and more stable than the previous DDPM-based diffusion \cite{simplespeech}. Furthermore, we also introduce an advanced transformer backbone from large language models, \textit{e.g.} time mixture-of-experts.

\end{itemize}



The rest of this paper is organized as follows: In Section 2, we motivate our study by introducing the background and related work. In Section~\ref{sec4:Proposed method}, we introduce the details of our proposed methods. The experimental setting, evaluation metrics, and results are presented from Section~\ref{sec:exp set} to Section~\ref{exp: experiments}. The study is concluded in Section~\ref{sec:conclusion}.


\begin{table*}[t]
\centering
\caption{Comparison of AR-based and NAR-based large-scale TTS methods.}
\label{tab1}
\begin{tabular}{|c|p{5cm}|p{5cm}|c|}
\hline
Method &
  Strengths &
  Potential Weaknesses &
  Representative Works \\ \hline
  \begin{tabular}[t]{@{}l@{}}  AR-based \\ large TTS \end{tabular} &
  \begin{tabular}[t]{@{}l@{}} \textcircled{1} Stand on the shoulder of LLMs \\   \textcircled{2} Diverse/expressive (sampling) \\ \textcircled{3} Simple data preparation \end{tabular} &
  \begin{tabular}[t]{@{}l@{}} \textcircled{1} Not stable/robust \\          \textcircled{2} Slow Inference\end{tabular} & \begin{tabular}[t]{@{}l@{}}
  VALL-E \cite{valle} \\ SPEAR-TTS \cite{speartts} \end{tabular} \\ \hline
\begin{tabular}[t]{@{}l@{}} NAR-based \\ large TTS \end{tabular} &
  \begin{tabular}[t]{@{}l@{}} \textcircled{1} Stable/Robust \\ 
 \textcircled{2} Fast inference \\           
 \textcircled{3} Control/Disentangle\end{tabular} &
  \begin{tabular}[t]{@{}l@{}} \textcircled{1} Over-smoothness (fidelity, prosody) \\ \textcircled{2} Less diversity \\ \textcircled{3} Complicated alignment process \end{tabular} & \begin{tabular}[t]{@{}l@{}} 
  NaturalSpeech 2 \cite{ns2} \\ NaturalSpeech 3 \cite{ns3} \end{tabular} \\ \hline
\end{tabular}
\end{table*}

\section{Related work and background} \label{sub: related work}
\subsection{Speech Tokenizer} \label{sec:related_work_tokenizer}
In this study, we define an \textit{audio/speech tokenizer} as a model that maps complex audio data into a latent space (latent representation), which can subsequently be reconstructed from this latent space back into audio. The latent representation may take the form of either discrete tokens or continuous vectors. Within the existing literature, the concept of an audio tokenizer has been extensively investigated \cite{yang2023hifi,encodec,soundstream,dac,make-an-audio}. These studies generally fall into two categories: Audio Codecs and Variance AutoEncoders (VAEs). Audio Codecs aim to quantize audio data into a set of discrete tokens, whereas VAEs are designed to model the mean and variance distributions of audio data.

Although audio codec models have been widely used in audio generation tasks, they face the problem of codebook collapse and need complex training loss combinations. Furthermore, the audio codec quantizes each speech frame into multiple tokens with RVQ, this will ensure high-quality audio reconstruction but will cause difficulty in autoregressive model generation \cite{valle,speartts} (error propagation and robust issues) due to the increased length of the token sequence. The VAE model or the continuous representations from audio codec are also can be used as the prediction target of generative models. For instance, previous works \cite{ns2,ditto-tts} follow the Latent Diffusion Model (LDM) \cite{ldm} to train a diffusion model to fit the distribution of continuous representations. However, developing a good VAE model also requires balancing the variance level with the reconstruction performance. A typical example is that data compressed with AE might reconstruct better than those with VAE, but the complexity retained in the compressed data distribution still poses challenges for the generative model. Thus, current VAE models in the audio and image generation fields imposes a slight Kullback-Leibler (KL) penalty towards a standard normal on the learned latent \cite{ldm,make-an-audio}. 
In this study, we propose to use scalar quantization to replace the Residual Vector Quantization (RVQ) in previous audio codec works, named SQ-Codec. We demonstrate that SQ-Codec not only achieves good reconstruction performance but also provides a finite, compact latent space, which is highly suitable for generative models.

\subsection{Large-scale text-to-speech}
Large-scale text-to-speech (TTS) research has made significant strides. These methods can be divided into two categories: autoregressive (AR) TTS based on large language models (LLMs) and non-autoregressive (NAR) TTS. AR-based TTS models got great attention, for instance, AudioLM \cite{borsos2023audiolm}, VALL-E \cite{valle}, SpearTTS \cite{speartts}, VoiceCraft \cite{voicecraft}, and UniAudio \cite{uniaudio} use audio codec (\textit{e.g.} Encodec \cite{encodec} or SoundStream \cite{soundstream}) for mapping speech to tokens, framing text-to-speech tasks as AR language modeling tasks, thus enabling zero-shot capabilities in the speech domain. AR-based TTS models are easy to train because they only rely on speech-transcription pairs data, but they also face the problem of slow generation speed and unstable generation quality. To improve its robustness, VALLE-R \cite{valle-r}, RALL-E \cite{rall-e} proposes to use more alignment information (\textit{e.g} phone duration) in the training stage, but such strategy also increases the difficulty of data preparation. 

The other type relies on NAR generative models, for instance, NaturalSpeech 2 \cite{ns2} proposes to model the continuous representation of the audio codec model by latent diffusion models. NaturalSpeech 2 is an NAR-based model, which has faster generation speed and better robustness than previous AR-based models. NaturalSpeech 3 \cite{ns3} proposes to encode different attributes (\textit{e.g.} content and prosody) into different codebooks, then applying discrete diffusion models \cite{diffsound,soundstorm,yang2024instructtts} on them. One of the drawbacks in NaturalSpeech 2/3 is that the system training needs fine-grained alignment information (\textit{e.g.} phone-level duration), which significantly hinders the use of large-scale in the wild data. Furthermore, the hard boundary alignment between phoneme and speech representation may result in unnatural prosody. Similarly, Voicebox \cite{voicebox} and Audiobox \cite{audiobox} propose a pre-training and fine-tuning strategy based on flow matching \cite{flow-matching}. HierSpeech ++ \cite{hierspeech++} proposes a hierarchical variational inference method. E3-TTS proposes to generate fixed-length audio based on text using diffusion models without any duration information, which significantly simplifies the data preparation for NAR TTS training. However, the fixed length generation also limits the flexibility of the model. Our recent publication SimpleSpeech \cite{simplespeech} proposes to use a sentence duration to control the length of audio, and we demonstrate that the sentence duration is a effective way for NAR TTS. In this study, we follow the line of our previous publication SimpleSpeech uses sentence duration to control the speech length. In addition, we investigate more strategies to obtain the sentence duration, refer to Section \ref{sec:sentence duration} to find the details. 

We notice that some concurrent works, SeedTTS \cite{seedtss}, DiTTo-TTS \cite{ditto-tts}, and E2TTS \cite{e2tts} also adopt a similar idea to use the sentence duration to control the generated speech length in an NAR TTS framework. We claim that our study is one of the earliest works to adopt this idea, and our proposed method is different from theirs.

\subsection{Diffusion Models}
Diffusion models \cite{ho2020denoising, sohl2015deep} have demonstrated great performance on a variety of generation tasks, including image \cite{ldm} and audio generation \cite{diffsound,make-an-audio,diffwave}. Many previous TTS systems that use diffusion models have been shown to produce high-fidelity speech that is comparable with state-of-the-art systems \cite{norespeech,wavegrad2,ns2,diffvoice,e3tts}. However, most of the previous diffusion-based TTS models rely on an additional duration modeling unit \cite{ns2,diffvoice} or carefully designed architecture \cite{e3tts}. In this study, we follow our previous publication, SimpleSpeech \cite{simplespeech}, to design a simple transformer-based diffusion model to generate speech without any alignment module. 
\begin{figure}[t] 
  \centering
  \includegraphics[width=0.8\linewidth, height=0.8\linewidth]{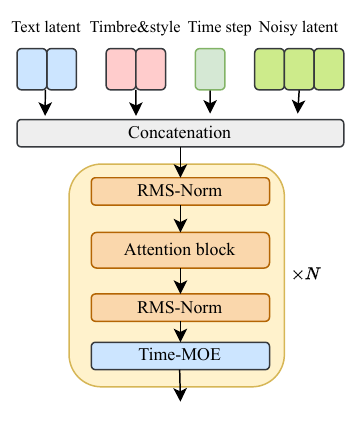}
  \caption{The structure of flow-based transformer diffusion.}
  \label{fig:dit}
  \vspace*{-\baselineskip}
\end{figure} 

\section{Proposed Method} \label{sec4:Proposed method}
In this part, we present the details of the proposed SimpleSpeech 2. 
The overall architecture of the SimpleSpeech 2 framework is demonstrated in Figure 1, which includes three parts: (1) off-the-shelf pre-trained text and speaker encoders to encode the text and speaker prompt without relying on speech domain-specific modeling, such as phoneme and duration; (2) A pre-trained high-quality audio tokenizer; (3) A LLAMA-style flow-based diffusion transformer. In the following, we first introduce the text encoder, speaker encoder, and SQ-Codec, and then give the details of flow-based scalar latent transformer diffusion models. 
\begin{figure}[t] 
  \centering
  \includegraphics[width=0.8\linewidth, height=0.6\linewidth]{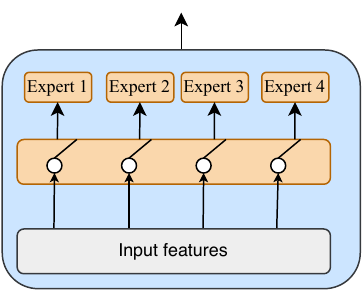}
  \caption{The structure of the proposed time mixture of experts (Time-MOE).}
  \label{fig:dit}
  \vspace*{-\baselineskip}
\end{figure} 
\subsection{Text Encoder} \label{text encoder}
This work explores to use the large-scale speech-only datasets to train a TTS system. Previous works use high-quality text-speech pair or their internal annotation tools to obtain text transcriptions, making TTS more complicated to implement. In this study, we propose to use the open available ASR ( Whisper-base model \cite{whisper}) model to get transcripts. Following SimpleSpeech \cite{simplespeech}, we use a pre-trained language model to directly extract the textual representations (\textit{e.g.} ByT5 \cite{btT5}) from text input without relying on speech domain-specific knowledge, such as phoneme. \\
\subsection{Speech Attribute Disentanglement Based Speaker Encoder} \label{speaker encoder}
Previous works directly extract speaker prompt representation from audio codec \cite{valle}, but such representation includes not only speaker timbre but also other acoustic information, such as recording condition, noise. Thus the synthesized quality is significantly influenced by the speaker prompt. For example, if the speaker prompt includes noise, the quality of synthesized speech will be poor. Inspired by the FACodec \cite{ns3}, which proposes to disentangle speech waveform into subspaces of content, prosody, timbre, and acoustic details, we use the pre-trained FACodec to extract the timbre representation. Specifically, we first use the timbre encoder of FACodec to extract a global embedding. Then we extract a fixed number of style vectors from the prosody encoder by using segment-based average pooling. To better show the prosody variation over time, we first split the prompt speech into three segments, and then we extract 3 style vectors from the prompt. Lastly, we concatenate the timbre and style vectors. 
\subsection{SQ-Codec} \label{sq-codec}
Recently, residual vector quantization (RVQ) based audio codec models have shown their advantages in audio compression and audio generation. Many audio generation models use an RVQ-based audio codec as the audio tokenizer (\textit{e.g.} extracting discrete tokens), then use a transformer model to predict the discrete tokens. Furthermore, the continuous representations extracted from the encoder of the audio codec can also be used as the training target of latent diffusion models, \textit{e.g.} NaturalSpeech 2 \cite{ns2}. However, current RVQ-based codec models have drawbacks: (1) training a good RVQ-based codec model needs a lot of tricks and complicated loss design \cite{dac}; (2) the reconstruction performance is positively correlated with codebook number, but too many codebooks also bring burden for LM-based model; (3) The distribution of continuous representations extracted from the encoder of audio codec is complex and infinite, which also increases the modeling difficulty of diffusion models.  In this study, we follow SimpleSpeech \cite{simplespeech} to use scalar quantization \cite{balle2016end,mentzer2023finite} to replace the residual vector quantization in audio codec models, which can be trained with the reconstruction loss and adversarial loss without any training tricks. We show that the scalar quantization effectively maps the complex speech signal into a finite and compact latent space, which is suitable for the diffusion model (refer to Section \ref{sub:good tokenizer} for more details). 

Similar to RVQ-based audio codec, our SQ-Codec also consists of an encoder, a decoder, and a scalar quantization. For any input speech $\boldsymbol{x}$, the encoder first transfers it as hidden representation, $\boldsymbol{h} \in \mathcal{R}^{T*d}$, where $T$ and $d$ denote the number of frames and the dimension of each vector. Then for any vector $\boldsymbol{h}_i$, we use a parameter-free scalar quantization module to quantize $\boldsymbol{h}_i$ into a fixed scalar space, this operation can be easily completed by following the formula:
\begin{equation}\label{codebook}
  \boldsymbol{h}_i = \text{Tanh}(\boldsymbol{h}_i), \quad
  \boldsymbol{s}_i = \text{Round}(\boldsymbol{h}_i*S)/S,
\end{equation}
where $S$ is a hyper-parameter that determines the scope of scalar space. We discuss the influence of hyper-parameter $S$ for the reconstruction performance and generation performance in Section \ref{sub:good tokenizer}. To get gradients through the rounding operation, we follow a straight-through estimator from VQ-VAE \cite{vqvae}. From formula 
\ref{codebook}, we can see that the SQ operation first uses a \textit{tanh} activation function to map the value of features into $[-1, 1]$, then it uses a \textit{round} operation to further scale the value of range into $\text{2*S+1}$ different numbers. We named such value domain as \textit{scalar latent space}. We find that previous works \cite{mentzer2023finite,yu2023language} also adopt scalar quantization for the image tokenizer. We claim that our implementation is different from theirs and we have different targets to build the codec: previous work tries to quantize image data into discrete representation, but we want to compress the search space of diffusion models. \\
\subsection{Flow-based Scalar Latent Transformer Diffusion Models}
Previous diffusion-based text-to-speech (TTS) methods typically model the speech data in a continuous space, \textit{e.g.} mel-spectrogram \cite{norespeech}, VAE \cite{diffvoice}, or the encoder output of audio codec \cite{ns2}. With the help of the powerful generative modeling ability of diffusion models, these works obtain good performance. In this study, we propose a new perspective on the modeling target of diffusion models: a compact, complete, and finite space is easier and more suitable for diffusion models. Specifically, we first pre-training a scalar quantization-based audio codec (SQ-Codec), as section \ref{sq-codec} introduced. Then we train a flow-based diffusion model that transfers the noise distribution into the scalar latent space. In the following,  we first give the preliminaries of flow-based diffusion models, and then describe how to use flow-based diffusion models to parameterize the problem.
\subsubsection{Flow-based Diffusion models}
In this study, we adopt a flow-based diffusion model \cite{flow-matching,gao2024lumina} for text-to-speech, which has been shown powerful generative ability in the image and video domain \cite{sd3,gao2024lumina}. Following previous work \cite{gao2024lumina}, we
will refer to it as flow matching in the following discussion. flow matching uses a simple strategy to corrupt data: it linearly interpolates between noise and data in a straight line. More specifically, given the representation of speech data $\rvx \sim p(\rvx)$ and Gaussian noise $\varepsilon \sim \mathcal{N}(0, \mathbf{I})$, an interpolation-based forward process
can be defined as:
\begin{equation}\label{flow interpolation1}
 \boldsymbol{x}_t = \alpha_t \boldsymbol{x} + \beta_t \boldsymbol{\epsilon}
\end{equation}
where $\alpha_0 =0 $, $\beta_0 = 1$, $\alpha_1 = 1 $, $\beta_1 = 0$, and $t \in [0,1]$. Similar to the diffusion schedule, we can set $\alpha_t$ and $\beta_t$ to follow the cosine schedule $\alpha_t = sin(\frac{\pi t}{2}), \beta_t = cos(\frac{\pi t}{2})$ or linear interpolation schedule $\alpha_t = t, \beta_t = (1-t)$. In this study, we adopt the linear interpolation schedule, so formula \ref{flow interpolation1} can be written as:
\begin{equation}\label{flow interpolation}
 \boldsymbol{x}_t = t \boldsymbol{x} + (1-t) \boldsymbol{\epsilon}
\end{equation}
where $t \in [0,1]$. This formulation shows a uniform transformation with constant velocity between data and noise.
The corresponding time-dependent velocity field can be written as
\begin{equation}
\begin{aligned}
  v_t(x_t) &= \hat{\alpha}_t \boldsymbol{x} + \hat{\beta}_t \boldsymbol{\epsilon} \\
          &= \boldsymbol{x} - \boldsymbol{\epsilon}
\end{aligned}
\end{equation}
where $\hat{\alpha}_t$ and $\hat{\beta}_t$ denote time derivative of $\alpha$ and $\beta$. Following \cite{flow-matching,gao2024lumina}, we consider the probability flow ordinary differential equation (ODE) with a velocity field: 
\begin{equation}
    \label{flow_ode}
d \rvx_t=\rvv_{\theta}(\rvx_t, t) d t,
\end{equation}
where the velocity $\rvv$ is parameterized by a neural network $\theta$. By solving the probability flow ODE backward in time from $\varepsilon \sim \mathcal{N}(0, \mathbf{I})$, we can generate samples and approximate the ground-truth data distribution $p(x)$. We refer to Eq.~\ref{flow_ode} as a flow-based generative model.
However, parameterizing $\rvv_{\theta}(\rvx_t, t)$ is computationally expensive. Previous works \cite{flow-matching} have shown that estimating a vector field $\rvu_t$ that generates a probability path between $p_0$ and $p_1$ is equivalent.
By solving this flow-based generative model from $t = 0$ to $t = 1$, we can transform noise into a data sample using the approximated velocity fields $\rvv_{\theta}(\rvx_t, t)$. During training, the flow-matching objective directly regresses the target velocity
\begin{equation} \label{cfm}
    \mathcal{L}_{F M} =  \min _\theta \mathbb{E}_{t, p_t(\rvx)}\|\rvv_\theta(\rvx, t)- \hat{\alpha}_t \boldsymbol{x} - \hat{\beta}_t \boldsymbol{\epsilon} \|^2,
\end{equation}
The formula \ref{cfm} is called conditional flow matching loss \cite{flow-matching}, it is similar to the noise prediction or score prediction losses in diffusion models. These properties minimize the trajectory curvature and connect the target data and noise on a straight line. It has been demonstrated that flow probabilistic models~\cite{ma2024sit,liu2022flow,esser2024scaling} can learn diverse data distribution in multiple domains, such as images \cite{gao2024lumina,sd3}. In this work, we compare the flow formulation to existing DDPM \cite{simplespeech} in text-to-speech and demonstrate its benefits. 

Furthermore, to make sure the final output belongs to the scalar latent space, we propose to use the scalar quantization (SQ) regularization to scale the final prediction.
\begin{equation}\label{s ldm}
   \boldsymbol{\hat{x}} = SQ(\theta(\boldsymbol{\epsilon},T,\boldsymbol{c})) 
\end{equation}
where $\theta$ denotes the flow-based transformer, $SQ$ denotes the scalar quantization operation in formula (\ref{codebook}). 
\subsubsection{Architecture}
As illustrated in figure \ref{fig:dit}, we describe the transformer architecture components in the following part. In previous work \cite{simplespeech}, we use a GPT-2 style \cite{gpt2} backbone. Inspired by the advanced LLMs architecture \cite{llama2}, we propose to build a more advanced transformer backbone for the text-to-speech task. The difference from the following parts: \\
(1) \textbf{Rotary Position Embedding (RoPE)}: RoPE \cite{rope} is used as the positional encoding. RoPE  can encode
relative positions, which is more suitable for TTS tasks than Absolute position encoding, \textit{e.g.} generating long speech during the inference. \\
(2) \textbf{RMSNorm}: we substitutes all
LayerNorm with RMSNorm to improve training stability \cite{llama2}. Moreover, it incorporates key-query normalization (KQ-Norm) before key-query dot product attention computation. The introduction of KQ-Norm aims to prevent loss divergence by eliminating extremely large values within attention logits. \\
(3) \textbf{Condition strategy}: In SimpleSpeech \cite{simplespeech}, we find that cross attention condition style \cite{ldm}, and in context condition (concatenate the conditional sequence and noisy latent sequence) are both effective. In this study, we follow the in-context condition strategy that join the sequences of the two modalities (condition features and noisy latent sequence) for the self-attention operation, such that both representations can work in their own space yet take the other one into account. \\
(4) \textbf{Time Mixture-Of-Experts}
As the previous part introduced, our flow-based diffusion models solve ODE backward in $t \in[0,1]$ and generate samples approximating the scalar latent distribution. We can find that 1) if the timesteps near $0$, the input $\mathbf{x}$ is highly noised, and the network of these steps mainly intends to capture the semantic information from conditional information (i.e., text and timbre information); 2) instead, when timesteps near $1$, the input $\boldsymbol{x}_t$ includes most of semantic information, where the model focus more on refining the details ( \textit{e.g.} improving the speech quality). To enhance the capability of flow-based generative models, following \cite{feng2023ernie}, we divide all the timesteps uniformly into $4$ blocks, in which each block consists of consecutive timesteps and is assigned to one denoising expert. Intuitively, when more experts are used, each block contains fewer timesteps, allowing each expert to better focus on learning the characteristics of the specific denoising steps assigned to it. 
\begin{figure*}[h]
    \centering
    \includegraphics[width=0.95\textwidth, height=0.35\textwidth]{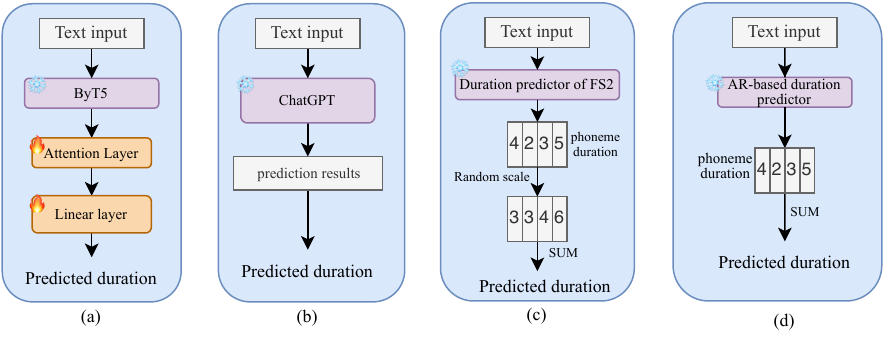}
    \vspace{-5mm}
    \caption{The overview of 4 different types of sentence duration predictor. (a) denotes that we use a pre-trained language model as the backbone to train a sentence duration predictor. (b) denotes that we use the in-context learning ability of ChatGPT to get sentence duration. (c) denotes that we make use of the pre-trained duration predictor from FastSpeech 2 to get sentence duration. (d) denotes that we utilize a pre-trained AR-based duration predictor.} 
    \label{fig:duration}
\end{figure*}
\subsection{Classifier-Free Guidance} \label{sec:cfc}
Classifier-free guidance (CFG) \cite{cfc} has been demonstrated as an effective way to enhance the generation quality in both image and audio domains \cite{sd3,yang2024instructtts}. CFG can be formulated as:
\begin{equation}\label{formula:cfc}
   \hat{p}(\boldsymbol{x}|\boldsymbol{y}) = p(\boldsymbol{x}) + \gamma (p(\boldsymbol{x}|\boldsymbol{y})-p(\boldsymbol{x}))
\end{equation}
where $\lambda$ denotes the guidance scale. CFC tries to model conditional distribution $p(\boldsymbol{x}|\boldsymbol{y})$ and unconditional distribution $p(\boldsymbol{x})$ in the training. In the inference stage, $\lambda=1$ denotes that we do not use classifier-free guidance, when $\lambda>1$ the model decreases the unconditional likelihood of the sample while increasing the conditional likelihood. In other words, classifier-free guidance
conducts this by decreasing the unconditional likelihood with a negative score term.  During the training stage, previous works try to mask the condition information of some samples (\textit{e.g.} set the training sample's text as empty with 10\% probability), so that these samples can be used to optimize unconditional distribution $p(\boldsymbol{x})$. 

Many prior studies \cite{valle,voicecraft} demonstrate that text transcriptions generated by an ASR system can be utilized to train a TTS system. Inevitably, the text transcription from the ASR system is not flawless, referred to as noisy or weak labels. Despite this, existing literature does not thoroughly explain why TTS systems can be effectively trained using such dataset. In this study, we show that including a small part of the noisy label in the training set is equivalent to introducing the CFC training strategy, thus we do not need to deliberately construct masked samples during training stage. In the following, we give the proof. 

Assume the condition is $\boldsymbol{y}$, the target is $\boldsymbol{x}$, and the generative model tries to model $p(\boldsymbol{x}|\boldsymbol{y})$. Consider the text $\boldsymbol{y}$ includes $n$ words, and $\boldsymbol{y}$ can be split into $n$ parts, and we have correspondence ($\boldsymbol{y}_i, \boldsymbol{x}_i$). Given that the TTS task tries to learn the mapping between $\boldsymbol{y}_i and \boldsymbol{x}_i$, $\boldsymbol{y}_i$. In general, we assume  $\boldsymbol{x}_j$ are mutually independent if $i \neq j$. The target of the generative model is to learn the distribution $p_{\theta}(\boldsymbol{x}|\boldsymbol{y})$. Based on Bayes formula, we have:
\begin{equation}\label{formula:bayes}
   \log p(\boldsymbol{x}|\boldsymbol{y}) = \log p(\boldsymbol{y}|\boldsymbol{x}) + \log p(\boldsymbol{x}) - \log p(\boldsymbol{y})
\end{equation}
We can discard the last term when we calculate the derivative of $x$.
\begin{equation}\label{formula:bayes-derivative}
   \nabla_x \log p(\boldsymbol{x}|\boldsymbol{y}) = \nabla_x \log p(\boldsymbol{y}|\boldsymbol{x}) + \nabla_x \log p(\boldsymbol{x}) 
\end{equation}
Based on the independence assumption, we can re-write $p(\boldsymbol{y}|\boldsymbol{x})$ as:
\begin{equation}\label{formula:bayes-derivative2}
\begin{aligned}
    p(\boldsymbol{y}|\boldsymbol{x}) &= p(y_1,y_2, ... y_n | x_1, x_2, ... x_n) \\
    &= \prod_{i=1}^{i=n} p(y_i|x_i)
\end{aligned}
\end{equation}
By combining equation \ref{formula:bayes-derivative2} and \ref{formula:bayes-derivative}, we have:
\begin{equation}\label{formula:bayes-derivative3}
   \nabla_x \log p(\boldsymbol{x}|\boldsymbol{y}) = \sum_i \nabla_x \log p(y_i|x_i) + \sum_{i} \nabla_x  \log p(x_i) 
\end{equation}
In a large-scale speech dataset, most of the words will occur many times. In some cases, the given input $y_i$ is not the ground truth label, we have $p(y_i|x_i) \to 0$, thus the first term in formula (\ref{formula:bayes-derivative3}) cannot contribute a positive gradient for our desired optimization process. In an extreme case, we set the text input as empty, so that each term of $p(y_i|x_i)=0$. It becomes an unconditional optimization problem:
\begin{equation}\label{formula:bayes-derivative4}
   \nabla_x \log p(\boldsymbol{x}|\boldsymbol{y}) =  \sum_{i} \nabla_x  \log p(x_i) 
\end{equation}
Note that the traditional CFC training strategy is to randomly choose mask samples during the training process, which means that each correct sample will be used to train the model. Instead, in our noisy label scene, these noisy labels are always wrong. Thus, the model just learns the unconditional distribution when noise labels are in the majority. 
\subsection{Sentence Duration} \label{sec:sentence duration}
In this study, we propose to use sentence duration to control the generated speech length in the inference stage. In our previous study, we use a sentence duration encoder \cite{simplespeech} to encode the duration information into the model. In this study, we further simply this module, we find that it is not necessary to encode the duration information. The sentence duration is only used to provide a coarse length guidance for the initialization of noisy latent.
To obtain the sentence duration, we propose the following potential strategies: \\
\textbf{ByT5 based sentence duration predictor} We propose to train a sentence duration prediction model with the help of a pre-trained large language model (ByT5 is used). As Figure \ref{fig:duration} (a) shows, we add several learnable attention and linear layers with the fixed ByT5 models, this model is asked to predict the duration. We train the model with MSE loss. \\
\textbf{Get sentence duration by using the context learning ability of ChatGPT} Following SimpleSpeech \cite{simplespeech}, we can use the in-context learning of ChatGPT to obtain the sentence duration. Specifically, we write a simple prompt for ChatGPT, and ask ChatGPT to predict a coarse sentence duration based on the length and the pronunciation of each words. Considering the ChatGPT is not stable, we call the ChatGPT 5 times for each text, and then get the average duration time. \\
\textbf{Obtaining sentence duration from teacher models} Previous NAR TTS models \cite{ns2,fastspeech2} has a phoneme duration predictor module, we can easily obtain the duration of each phoneme, and then we can sum them to get the final sentence duration. We use the phoneme duration predictor from open sourced FastSpeech 2 model \footnote{https://github.com/ming024/FastSpeech2}. To increase the diversity, we add a \textit{scale factor} for each phoneme duration, the scale factor is a random value, which belongs to $[0.9, 1.3]$. Note that we only use this module to obtain a coarse sentence duration, which does not introduce additional complexity for our SimpleSpeech 2. \\
\textbf{Training a AR-based duration model} We can also train a shallow AR-based duration predictor, that predicts the phoneme or word duration in an autoregressive style. In this study, we use the open-sourced phoneme predictor \footnote{https://github.com/ex3ndr/supervoice-gpt} as Fig. \ref{fig:duration} shows. Similarly, such module is used to obtain a coarse sentence duration, it does not introduce additional complexity for SimpleSpeech 2. \\
In Section \ref{exp:duration}, we give detailed experimental analysis for each way and find that the ByT5-based sentence duration predictor shows the best performance. 

\section{Experimental Setup} \label{sec:exp set}
\subsection{Dataset} \label{dataset}
\noindent \textbf{Training data} In this study, we train two distinct models, SimpleSpeech 2-EN and SimpleSpeech 2-MUL.
For SimpleSpeech 2-EN, we create a dataset with only 7k hours of unlabeled English speech data sampled from the Multilingual LibriSpeech (MLS) dataset \cite{mls}. For each original audio, we randomly crop 2-10 seconds if the length is larger than 10 seconds. Although the MLS dataset includes text labels, we opt not to use these in order to explore the potential of a speech-only dataset for TTS development. For all of the speech samples, we re-sampling it into 16k Hz. 
Given the predominant focus of previous research on English datasets, we designate SimpleSpeech 2-EN as our primary experiment and benchmark its performance against existing methods. \\
For SimpleSpeech 2-MUL, we enrich the dataset with an additional 4k hours Chinese dataset from WenetSpeech \cite{zhang2022wenetspeech}, which we integrate with the 7k hours English data from MLS. Leveraging the multilingual capabilities of the ByT5 model, we concurrently train on both English and Chinese speech. \\
\textbf{Evaluation data} 
Most prior studies train their models on the LibriLight \cite{librilight} or MLS \cite{mls} datasets, and then evaluate performance using the LibriSpeech test set \cite{libritts}. One potential drawback is that both LibriSpeech and LibriLight are sourced from the same LibriVox \cite{kearns2014librivox} audiobooks dataset, which may not fully represent the diversity of real-world audio environments. In this study, we propose to evaluate the model using test sets derived from various other sources. We have constructed a comprehensive test set that includes LibriTTS \cite{libritts}, VCTK, CommonVoice \cite{ardila2019common}, RAVDESS \cite{livingstone2018ryerson}, and SwitchBoard \cite{godfrey1992switchboard}. Both LibriTTS and VCTK are widely used in academic research and offer high-quality speech data. SwitchBoard, being a conversational test set, and RAVDESS, which provides a range of emotional speech samples, can validate the model’s ability to handle conversational dynamics and emotional variance, respectively.  \\
\subsection{Model training and inference} \label{subsec:model training}
\noindent \textbf{SQ-Codec training} For the SQ-Codec model, we train it on the LibriTTS \cite{zen2019libritts} dataset. The sample rate is set as 16k Hz. Following \cite{simplespeech}, we default use $S=9$ and $d=32$ for all of the experiments. We also train different SQ-Codec models by changing the hyper-parameter $S$ and $d$, refer to Section \ref{sub:good tokenizer} for more details. During the training, we set the learning rate as 2e-3. We used Adam optimizer for all of the experiments. We train the SQ-Codec models with 200k steps. \\
\textbf{SimpleSpeech 2 training} For the SimpleSpeech 2 model training, we freeze the SQ-Codec, text encoder, and speaker encoder. We only update the parameters of the transformer diffusion part. We train it by using Adam optimizer with the learning rate of $1e-4$. We use a cosine learning rate scheduler with a
warmup of 1K steps. The model is trained with 400k steps. For the transformer, we set 16 attention layers, the hidden size is 768, and the transformer head is 32. \\
\textbf{Inference} To synthesize speech, we first input the text into the proposed sentence duration module, which can be used to determine the total length of the predicted speech by controlling the length of the noisy latent sequence. Then we input the text and speaker prompt into the SimpleSpeech 2 to generate the predicted latent of SQ-codec. Lastly, we get the predicted waveform by inputting the predicted latent representation into the SQ-Codec decoder. \\
\subsection{Baselines}
\noindent \textbf{TTS Baselines} We compare the proposed model with state-of-the-art autoregressive TTS models: 
\begin{itemize}
\item VALL-E \cite{valle}: we reproduce it on the same dataset with us.
\item VALL-EX \cite{vallex}: we use the open-source version from Git Hub.\footnote{https://github.com/Plachtaa/VALL-E-X} 
\item VoiceCraft \cite{voicecraft}: we use the official open-source checkpoint from GitHub. \footnote{https://github.com/jasonppy/VoiceCraft}
\item ARDiT \cite{liu2024autoregressive}: We compared with their demo samples.
\item ChatTTS: A popular open-sourced AR-based TTS. \footnote{https://github.com/2noise/ChatTTS}
\end{itemize}
We also compare the proposed model with Non-autoregressive TTS models:
\begin{itemize}
\item NaturalSpeech 2 \cite{ns2}: we reproduce it on the same dataset with us.
\item NaturalSpeech 3 \cite{ns3}: we get the samples from the authors. 
\item HierSpeech++ \cite{hierspeech++}: we use the official open-source checkpoint from GitHub.\footnote{https://github.com/PolyAI-LDN/pheme}
\item E3TTS \cite{e3tts}: We compared with their demo samples.
\item DiTTo-TTS \cite{ditto-tts}: We compared it with their demo samples.
\end{itemize}
\textbf{Speech Tokenizer Baselines} For audio codec models, we compare the proposed SQ-Codec model with state-of-the-art models, includes: including Encodec \cite{encodec}, DAC \cite{dac}, HiFi-Codec \cite{yang2023hifi}, Soundstream \cite{soundstream}, and VAE \cite{stableaudio}. For EnCodec, DAC, and HiFi-Codec, we use their official checkpoints for inference. For SoundStream, we reproduce it based on the open-source repository \footnote{https://github.com/yangdongchao/AcademiCodec}. For VAE, we follow the open source repository of Stable Audio \cite{stableaudio} \footnote{https://github.com/Stability-AI/stable-audio-tools}. For both SoundStream and VAE, we set their latent dimension as the same as SQ-Codec and trained them with the same dataset as SQ-Codec.  \\
\subsection{Evaluation metrics}
\subsubsection{Speech Tokenizer Evaluation} For audio tokenizer models, we use the following metrics: Perceptual Evaluation of Speech Quality (PESQ), Short-Time Objective Intelligibility (STOI), and structural similarity index measure (SSIM). \\
\subsubsection{TTS Evaluation} For the TTS task, we assess the performance from 4 aspects: intelligibility, speaker similarity, speech quality, and generation speed. Furthermore, we also conduct subjective evaluations for TTS.  \\
\textbf{Intelligibility} Word Error Rate (WER) is widely used to assess the intelligibility of generated speech. However, previous work \cite{seedtss} suggests that WER may not serve as a definitive criterion; a lower WER often indicates that the model produces more ‘standardized’ speech, which is easier for ASR systems to recognize. To provide a more comprehensive evaluation, we propose using multiple ASR systems to assess the generated speech, employing the mean WER as the metric for comparison. In this study, we utilize the Whisper-large model and the Hubert-large-based ASR system to evaluate the robustness. \\
\textbf{Speaker Similarity} Similarly, previous works use a pre-trained speaker model to extract speaker embedding, and then calculate the speaker similarity score (SIM). However, we empirically observe that using different speaker embedding models results in different SIM scores. Thus, we propose to use multiple speaker embedding models as the 'referee', and use the average score as the final metric. More specifically, we use the pre-trained WavLM-TDNN model and the 1st layer of XLSR-53 \cite{conneau2020unsupervised} to extract speaker embedding. \\
\textbf{Speech quality (Perception score)} DNSMOS \cite{dnsmos} and Mel-Cepstral Distortion (MCD) are used to measure speech quality. \\
\textbf{Generation speed} Following previous work, we use RTF real-time factor (RTF) to measure the efficiency:
\begin{equation}
 RTF = \frac{Cost\_Time}{Duration}
\end{equation}
\noindent \textbf{Subjective Evaluation} For subjective metrics, we hire 15 listeners to conduct speech naturalness/quality (MOS) and speaker similarity (SMOS) tests. For the MOS test, we ask listeners to focus on whether the speech is realistic, the speech's naturalness, clearness, and prosody. For the SMOS test, we ask listeners to judge whether the generated speech is similar to the reference speech in terms of timbre, emotion, and prosody. \\

 \begin{table}[t]
    \centering
    \small
    \caption{Reconstruction Performance comparison between open-sourced audio tokenizers and the proposed SQ-Codec. B denotes the Bitrate (kbps). H denotes the Hop-size. * denotes that the model is our reproduced due to official checkpoints are not available.}
    \vspace{2mm}
    \scalebox{0.92}{
    \begin{tabular}{lccc|ccc}
    \toprule
    Model  &Size (M)  & H & B  & PESQ  & STOI & SSIM  \\
    \midrule
    Encodec & 14   & 75 & 12 & 3.76  & 0.90 & 0.72       \\
    DAC  & 70  & 50 & 8 & 3.99 & 0.95 & 0.85    \\
    HiFiCodec  & 60 & 50 &  2  & 3.24 &  0.88  & 0.72    \\
    SoundStream* & 14   & 50 & 4 & 3.25  & 0.90 & 0.77       \\
    VAE*  & 15 & 50 &  -  & 3.93 &  0.94  & 0.86    \\
    \midrule
    SQ-Codec (Ours)  & 5  & 50 & 8  & \textbf{4.16} & \textbf{0.95 }& \textbf{0.86} \\
    \bottomrule 
    \end{tabular}
    \label{tab:all-codec-comparison} }
\end{table}
\section{Experiments} \label{exp: experiments}
\subsection{The performance of Speech Tokenizer}
As we discussed in previous sections, speech tokenizers play an important role in the proposed TTS system: (1) Completeness: its reconstruction performance determines the upper limit of speech generation; (2) Compactness: which determines the search space of the generative model. We first compare the reconstruction performance of our proposed SQ-Codec model to other state-of-the-art audio tokenizers. Table \ref{tab:all-codec-comparison} shows the results, we can see that our proposed SQ-Codec has a good reconstruction performance, which lays the foundation for high-quality speech generation. We have the following findings: (1) Compared with previous audio codec models, such as Encodec and Soundstream, our SQ-Codec has better reconstruction performance. Especially, our SQ-Codec has better reconstruction performance than VAE. As we discussed in Section \ref{sec:related_work_tokenizer}, VAE needs to force the data distribution belongs to normal distribution, which may limit the reconstruction performance. (2) We note that SQ-Codec is more lightweight than other models, which further improves the generation efficiency. 
Furthermore, we also conduct an ablation study to explore the influence of two hyper-parameters ($S$ and $d$) in SQ-Codec. Specifically, we evaluate the reconstruction performance and the generation performance when the corresponding audio tokenizer is used. Please refer to Section \ref{sub:good tokenizer} to see our findings.

\begin{table*}[!h]
    \centering
	\caption{The comparison with previous zero-shot TTS models. MOS and SMOS are presented with 95\% confidence intervals. $\spadesuit$ means the reproduced results. $\heartsuit$ means the results are inferred from official checkpoints. $\bigstar$ means the results from the authors. \textbf{Bold} for the best result and \underline{underline} for the second-best result}
    \label{table:tts}
    \begin{tabular}{ l >{\centering}p{1.2cm} >{\centering}p{1.6cm} c c c c c c c c}
        \toprule
        \multicolumn{1}{c}{} & \multicolumn{2}{c}{MUSHRA ($\uparrow$)} & \multicolumn{3}{c}{WER ($\downarrow$)} & \multicolumn{3}{c}{SECS ($\uparrow$)}  & \multicolumn{1}{c}{Speed ($\downarrow$)} & \multicolumn{1}{c}{Perception ($\uparrow$)} \\
        \cmidrule(r){2-3} \cmidrule(r){4-6} \cmidrule(r){7-9} \cmidrule(r){10-10} \cmidrule(r){11-11}
        {Model Name}  & MOS   & SMOS  & Whisper &Hubert &AVG     & WavLM  & Wav2Vec &AVG   & RTF  & DNSMOS      \\
        \midrule
        Ground Truth   & 3.96$\pm$0.18  & 4.13$\pm$0.11  & 5.06 & 9.7 & 7.37 & 0.94          & 0.97 & 0.96    & -  & 3.61        \\
        \midrule
        VoiceCraft \cite{voicecraft} $\heartsuit$  & 3.53$\pm$0.36 & 3.70$\pm$0.40  & 18.9 & 25.0 & 21.9  & 0.86 & 0.91
          & 0.89    & 1.92        & 3.55          \\
        VALL-E \cite{valle} $\spadesuit$   & 3.43$\pm$0.25 & 3.60$\pm$0.12  & 17.4 & 25.0 & 23.8  & 0.85 & 0.93 & 0.89    & 5.4        & 3.44          \\
        VALL-EX \cite{vallex} $\heartsuit$   & 3.60$\pm$0.19 & 3.68$\pm$0.33  & 19.3 & 28.4 & 23.8  & 0.88 & 0.94 & 0.91    & 10.2  & 3.73          \\
        X-TTS \cite{xtts} $\heartsuit$   & 4.00$\pm$0.20 & \textbf{3.97$\pm$0.14}  & 5.12 & 12.8 & 8.96  & 0.89 & 0.92 & 0.91    & 1.21    & 3.66          \\
        ChatTTS $\heartsuit$   & 3.90$\pm$0.17 & -  & 5.9 & 12.9 & 9.4  & - & - & -    & 1.36    & 3.68          \\
        \midrule
        HierSpeech++ \cite{hierspeech++} $\heartsuit$  & 3.85$\pm$0.16   & 3.81$\pm$0.14   & \textbf{3.66}   & \textbf{8.0}   & \textbf{5.83}  & 0.88 & 0.95 & 0.91 & 0.57 & 3.71         \\
        NaturalSpeech 2 \cite{ns2} $\spadesuit$ & 3.58$\pm$0.17   & 3.65$\pm$0.18  & 5.58   & 12.5  & 9.04  & 0.85 & 0.93 & 0.89 & 2.7 & 3.36         \\
        NaturalSpeech 3 \cite{ns3} $\bigstar$  & 3.96$\pm$0.27  & 3.89$\pm$0.24  & 6.3  & 11.6   & 8.96  & \textbf{0.91}  & 0.95 & 0.93 & 0.73 & 3.48          \\
        SimpleSpeech \cite{simplespeech} $\heartsuit$  & 3.97$\pm$0.21   & 3.75$\pm$0.33  & 9.3   & 17.8  & 13.5  & 0.87 & 0.95 & 0.91 & 1.6 & 3.79         \\
        \midrule
        SimpleSpeech 2 (Ours)   & \textbf{4.28$\pm$0.12}  & \underline{3.90$\pm$0.26}   & \underline{5.10}   & \underline{10.0}  & \underline{7.55}  & \underline{0.89} & \textbf{0.97} & \textbf{0.93} & \textbf{0.25} & \textbf{3.85}         \\
        \bottomrule 
    \end{tabular}\\
\end{table*}

\subsection{Comparison with Previous Large-scale TTS Baselines}
In this section, we conduct experiments to evaluate the performance of zero-shot speech synthesis. Note that we compare our SimpleSpeech 2-EN version with other baselines in the English dataset. Table \ref{table:tts} and \ref{tab:small-scale} show the results. In the following, we will analyze the results from three aspects: Speech quality, Speaker Similarity, Robustness, and Generation Speed. \\
\textbf{Speech quality} Speech quality is an important factor in evaluating a TTS system, previous works may result in poor speech quality due to suboptimal audio tokenizer and generative strategies. In this study, we adopt a more advanced audio tokenizer and generative model, which significantly improves the speech quality. From the MOS and DNSMOS scores, we can see the generated samples by SimpleSpeech 2 even surpass the ground truth speech. \\
\textbf{Speaker Similarity} Nowadays, voice cloning ability has been viewed as a basic metric to access the zero-shot TTS system. Previous works \cite{ns3,voicebox,voicecraft} have shown that scaling the speech data can improve the zero-shot cloning ability. Most of them are both trained on large-scale datasets (\textit{e.g.} 60k hours). Although our SimpleSpeech 2 only trained on 7k hours of data, we can see that it also shows comparable performance to baselines trained with much more data. \\
\textbf{Robustness} As we discussed in Section \ref{sub: related work}, AR-based TTS systems usually face the robustness problem because their training does not rely on phoneme-level duration alignment data. In this study, our proposed model also does not rely on phoneme-level alignment data. From Table \ref{table:tts}, we can see that SimpleSpeech 2 shows better robustness than previous AR-based works, \textit{e.g.} SimpleSpeech 2 is more robust than VALL-EX, ChatTTS, VoiceCraft in our English evaluation set. Compared with phoneme duration-driven NAR systems, our method is still behind some NAR-based models, such as HierSpeech++. We believe that the gap can be eliminated by scaling data. The concurrent work, SeedTTS \cite{seedtss}, also uses a similar strategy to us, they show that the system can be more robust after training with large-scale data. \\
\textbf{Generation Speed} In general, AR-based large TTS models will cost a lot of time in inference because they generate speech tokens one by one. In this study, our model can generate scalar latent features in an NAR way, then the decoder of SQ-Codec is used to recover waveform from the generated feature. From Table \ref{table:tts}, we can see that our proposed model is better than AR-based models in terms of RFT. Compared with other NAR TTS models, our proposed method needs fewer inference steps without any diffusion distillation. For instance, E3TTS \cite{e3tts} proposes to generate a waveform, it needs 1000 diffusion steps. NaturalSpeech 2 also uses 150 steps to ensure sufficient generation quality. Instead, our SimpleSpeech 2 is based on flow-based diffusion models, it can generate high-quality speech with 25 steps. As a contrast, our previous work, SimpleSpeech uses 100 diffusion steps. 
\begin{table}[t]
    \centering
    \small
    \vspace{-2mm}
    \caption{The small-scale Subjective Evaluation. }
    \vspace{2mm}
    \begin{tabular}{lcccc}
    \toprule
    Model   & E3TTS \cite{e3tts}  & ARDiT  & DiTTo-TTS \cite{ditto-tts} & Ours \\
    \midrule
      CMOS & -0.57 & -0.5 &  -1.3 & \textbf{0.0}   \\
      SMOS & 4.08$\pm$0.4 & 4.0$\pm$0.3 & 3.85$\pm$0.4 & \textbf{4.08$\pm$0.2}  \\
    \bottomrule 
    \end{tabular}
    \vspace{-3mm}
    \label{tab:small-scale} 
\end{table}

 \begin{table}[t]
    \centering
    \small
    \caption{The influence of audio tokenizer for generation performance.}
    \vspace{2mm}
    \scalebox{0.92}{
    \begin{tabular}{lcccc}
    \toprule
    Model    & WER ($\downarrow$)  & SIM ($\uparrow$) & DNSMOS ($\uparrow$) &MCD ($\downarrow$)  \\
    \midrule
    SoundStream  & 9.7  & 0.91 & 3.69 & 6.57      \\
    VAE    & 10.5 & 0.91 & 3.77  & 6.62   \\
    SQ-Codec  & \textbf{7.5} & \textbf{0.93} & \textbf{3.85} & \textbf{6.42} \\
    \bottomrule 
    \end{tabular}
    \label{tab:ablation-features}}
\end{table}
\subsection{Discussion: what is a good audio tokenizer?} \label{sub:good tokenizer}
In this section, we explore the influence of audio tokenizers on large-scale TTS systems. Initially, we replace our proposed SQ-Codec with other audio tokenizers, such as VAE and SoundStream. The results are presented in Table \ref{tab:ablation-features}. 
Subsequently, we investigate the impact of the search space of SQ-Codec on speech generation. Specifically, we set different hyper-parameters,  $S$ and $d$ , to train new SQ-Codec models. These models are then utilized as audio tokenizers for speech synthesis. Our findings are as follows: \\
(1) The reconstruction performance of the audio tokenizer determines the upper limit of generated speech quality.For example, Table \ref{tab:all-codec-comparison} shows that our SQ-Codec achieves the best reconstruction performance, while our reproduced version of SoundStream performs less effectively. Consequently, using SQ-Codec as the audio tokenizer results in the highest DNSMOS score for the generation model, whereas using SoundStream yields inferior outcomes. \\
(2) The proposed scalar latent space is more readily modeled by generative model. In Table \ref{tab:ablation-features}, we set the latent dimension of VAE, SoundStream, and SQ-Codec as 32. The difference is that SoundStream and VAE tried to compress the speech data into a complex continuous distribution, and SQ-Codec tried to compress the speech data into a finite distribution: the value range of each pixel in the scalar latent space belongs to a discrete uniform distribution. Instead, the range of SoundStream and VAE is infinite. Consequently, using SQ-Codec as the speech tokenizer results in superior performance. \\
(3) The Compactness of audio tokenizers significantly influences the performance of generative models. As Table \ref{tab:ablation-sq-codec} shows, when we increase the latent dimension of SQ-Codec to 50, its reconstruction performance obtains improvement, but the performance of the generation model significantly decreases. Similarly, if we decrease the latent dimension and scalar value, the reconstruction performance will drop, and a poor audio tokenizer also influences the generation performance. In our study, to balance the completeness and compactness, we suggest to use $S=9$ and $d=32$. \\
In summary, an effective tokenizer should compress speech data into a latent space that is both compact and complete, ensuring good reconstruction performance and ease of modeling by generative models.
 \begin{table*}[t]
    \centering
    \small
    \caption{The influence of different parameters of SQ-Codec for reconstruction and generation. }
    \vspace{2mm}
    \begin{tabular}{ l >{\centering}p{1.2cm} >{\centering}p{1.6cm} c c c c c c c c}
        \toprule
        \multicolumn{1}{c}{} & \multicolumn{1}{c}{} & \multicolumn{3}{c}{Reconstruction Performance}  & \multicolumn{4}{c}{Generation Performance }   \\
        \cmidrule(r){3-5} \cmidrule(r){6-9} 
        {Scalar value} & Latent    & PESQ ($\uparrow$)   & STOI ($\uparrow$) & SSIM ($\uparrow$) & WER ($\downarrow$) & SIM ($\uparrow$) & MCD ($\downarrow$)  & DNSMOS  ($\uparrow$)        \\
        \midrule

    9 & 32   & 4.16  & 0.95 & 0.86 & \textbf{7.5}  & \textbf{0.93} & 6.42 & \textbf{3.85}       \\
    9  & 20  & 3.83 & 0.934 & 0.79 & 8.6 & 0.92 & 5.63 & 3.76      \\
    9  & 10  & 3.39  & 0.89 & 0.74  & 11.5  & 0.91 & 6.39 & 3.81        \\
    9  & 50  & \textbf{4.23} &  \textbf{0.97}  & \textbf{0.86} & 18.9  & 0.92 & 6.88 & 3.82      \\
    \midrule
    3 & 32   & 3.90  & 0.94 & 0.80  & 13.9  & 0.93 & \textbf{6.25} & 3.84      \\
    \midrule
    5  & 32  & 4.01 &  0.95  & 0.81  & 9.1  & 0.92 & 6.66 & 3.84       \\
    \bottomrule 
    \end{tabular}
    \label{tab:ablation-sq-codec}
\end{table*}
\subsection{Discussion: why ASR transcription can be used as a label to train a good TTS system?}
Many previous studies \cite{valle,uniaudio} have demonstrated that text transcriptions from ASR systems can be effectively used to train TTS systems. However, these transcriptions are inevitably imperfect and are often referred to as noisy or weak labels. In Section \ref{sec:cfc}, we provide a theoretical analysis showing that including a small number of noisy labels in a large-scale dataset is analogous to employing classifier-free guidance training for model optimization. Generally, most ASR systems perform well, achieving a WER of less than 10\%. Therefore, we can utilize these ASR transcriptions to train our TTS system. In this study, we employ a state-of-the-art (SOTA) open-sourced ASR model, Whisper \cite{whisper}, which has been widely validated.

\subsection{Ablation study: the influence of sentence duration predictor} \label{exp:duration}
\begin{figure}[t] 
  \centering
  \includegraphics[width=\linewidth]{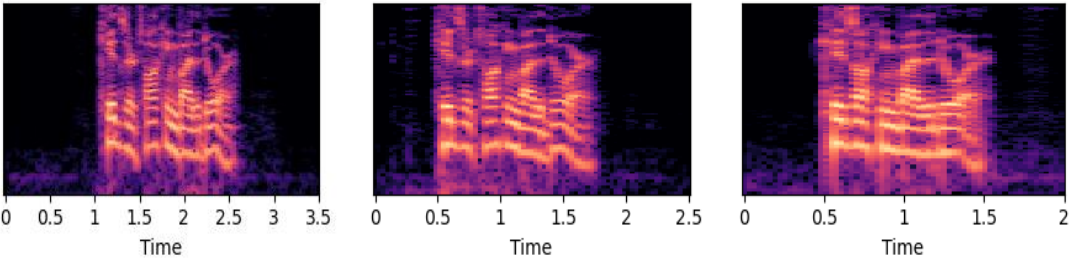}
  \caption{The mel-spectrogram visualization for speech synthesized from the same content and audio prompt, but using different sentence durations.}
  \label{fig:du-vis}
  \vspace*{-\baselineskip}
\end{figure}

\begin{table}[t]
    \centering
    \small
    \caption{The performance comparison when we use different sentence duration predictors for the TTS system.}
    \vspace{2mm}
    \scalebox{0.92}{
    \begin{tabular}{lcccc}
    \toprule
    Duration predictor    & WER ($\downarrow$)   & SIM ($\uparrow$) & MCD ($\downarrow$) & DNSMOS ($\uparrow$)   \\
    \midrule
    ByT5 based   & \textbf{7.55}  & 0.933 & 6.42 & \textbf{3.85}       \\
    FS2-based  & 7.96 & \textbf{0.937} & \textbf{6.36} & 3.84    \\
    AR-based  & 8.64 &  0.925  & 6.40 & 3.81     \\
    ChatGPT  & 9.61 & 0.929  & 6.38 & 3.84 \\
    \midrule
    Ground Truth  & 7.40 & 0.930  & 6.24 & 3.85 \\
    \bottomrule 
    \end{tabular}
    \label{tab:ablation-duration}}
\end{table}
In this part, we explore which type of sentence duration is better. From Table \ref{tab:ablation-duration}, we observe that (1) Four types of sentence duration models are effective. (2) Compared to specialized models, ChatGPT performs poorly in predicting sentence duration. One reason for this is the instability of ChatGPT’s predictions, which sometimes produce extreme cases, such as overly short predicted durations. Although increasing the number of demonstration samples in the prompt and averaging multiple runs can mitigate this issue, it also significantly raises inference costs. (3) ByT5-based predictors and the duration predictor of FastSpeech 2 \cite{fastspeech2} demonstrate greater stability than AR-based phoneme duration predictors.  Considering both cost and simplicity, we recommend using the ByT5-based sentence duration predictor as default choice. Fig. \ref{fig:du-vis} illustrates how SimpleSpeech 2 can synthesize diverse prosodies using different sentence durations.
\subsection{Ablation study: Architecture and Formulation}
In this section, we validate the effectiveness of our architecture design, specifically the Time-MoE module, and our flow-based diffusion formulation. The results, as presented in Table \ref{tab:ablation-arche}, demonstrate the efficacy of the Time-MoE design and the flow-based diffusion approach.
 \begin{table}[t]
    \centering
    \small
    \caption{Ablation studies on architecture and formulation.}
    \vspace{2mm}
    \scalebox{0.92}{
    \begin{tabular}{lcccc}
    \toprule
    Setting     & WER  & SIM & MCD  & DNSMOS  \\
    \midrule
    SimpleSpeech 2  & \textbf{7.5}  & \textbf{0.93} & 
    \textbf{6.42} & \textbf{3.85}       \\
    \midrule
    w/o Time-MoE    & 8.2 &  0.92  & 6.67 & 3.80    \\
    DDPM formulation  & 9.0 & 0.92  & 6.62 & 3.82 \\
    \bottomrule 
    \end{tabular}
    \label{tab:ablation-arche}}
\end{table}
                    
\subsection{The influence of classifier free guidance}
We also explore the effectiveness of classifier-free guidance (CFG), as documented in Table \ref{tab:cfc}, which presents the experimental results. These findings reveal that the CFC strategy significantly influences generation performance. For example, omitting the CFG strategy (by setting $\lambda=1$) results in decreased performance. Furthermore, different parameter configuration also influences the performance. In this study, we default set $\lambda=5$ for experiments.

\begin{table}
    \centering
    \small
    \caption{The influence of classifier-free guidance.}
    \vspace{2mm}
    \scalebox{0.92}{
    \begin{tabular}{lccc}
    \toprule
    CFC Scale ($\lambda$)    & WER ($\downarrow$)  & SIM ($\uparrow$)  & DNSMOS ($\uparrow$)   \\
    \midrule
    1   & 50.4  & 0.910 & 3.58       \\
    2   & 16.7 &  {0.935} & 3.74    \\
    3  & 10.4 & {0.935}  & 3.85  \\
    4  & 7.91 & 0.929  & 3.83 \\
    5  & {7.5} & 0.933  & {3.85} \\
    6  & 8.5 & 0.925  & 3.81 \\
    7  & 8.0 & 0.925  & 3.80 \\
    \bottomrule 
    \end{tabular}
    \label{tab:cfc}}
\end{table}
\subsection{The ablation study of diffusion steps}
Table \ref{tab:step} shows the generation performance when we use different diffusion steps. We can see that using more diffusion steps can improve the generation performance, but also sacrificing the inference time. In our study, we find that 25 diffusion step is good enough to obtain good generation performance.
\begin{table}[t]
    \centering
    \small
    \caption{The influence of diffusion step.}
    \vspace{2mm}
    \scalebox{0.92}{
    \begin{tabular}{lcccc}
    \toprule
    Diffusion Step  & RTF ($\downarrow$)   & WER ($\downarrow$)  & SIM ($\uparrow$)  ($\downarrow$) & DNSMOS ($\uparrow$)   \\
    \midrule
    5   & 0.12  & 13.85 & 0.91 & 3.76       \\
    10   & 0.15 & 10.25  & 0.92 & 3.81    \\
    15  & 0.19 & 10.30 & 0.92 & 3.83  \\
    20  & 0.21 & 11.20 & 0.93 & 3.84 \\
    25  & 0.25 & 7.55 & 0.93 & 3.85 \\
    50  & 0.42 & 7.43 & 0.93 & 3.85 \\
    \bottomrule 
    \end{tabular}
    \label{tab:step}}
\end{table} 
\subsection{SimpleSpeech 2-MUL: Extending to multilingual TTS}
\begin{table}[t]
    \centering
    \small
    \caption{The performance of SimpleSpeech-MUL in Chinese dataset.}
    \vspace{2mm}
    \scalebox{0.92}{
    \begin{tabular}{lcccc}
    \toprule
    Model    & WER ($\downarrow$)  & SIM ($\uparrow$) & MCD ($\downarrow$) & DNSMOS ($\uparrow$)  \\
    \midrule
    Ours    & 13.98 & 0.94 & \textbf{5.80}  & \textbf{3.65}    \\
    ChatTTS & \textbf{8.88} & - &  6.17 & 3.46 \\
    \bottomrule 
    \end{tabular}
    \label{tab:mul}}
\end{table}
Thanks to the ByT5 text encoder, SimpleSpeech 2 supports multilingual text-to-speech synthesis. Specifically, we trained SimpleSpeech 2-MUL using Chinese and English datasets. For inference, we selected 30 sentences from the AISHELL 3 dataset to evaluate Chinese TTS performance. We compared our model with ChatTTS, which is trained on a large-scale Chinese dataset. As shown in Table \ref{tab:mul}, our model achieves comparable performance to ChatTTS. However, we recognize that there is still room for improvement in the robustness of our model on the Chinese dataset. One potential factor is the direct input of Chinese words into the ByT5 model, which complicates modeling due to the prevalence of polyphones in Chinese. We hypothesize that expanding to a larger dataset could enhance pronunciation accuracy. Additionally, our model demonstrates the ability to produce natural code-switched speech; for more details, please visit our demo page.
\section{Conclusion} \label{sec:conclusion}
In this study, we introduce SimpleSpeech 2, an enhanced TTS system building upon our previous work. SimpleSpeech 2 incorporates advanced model structures, such as Time Mixture-of-Experts, and employs flow-based scalar diffusion techniques. We demonstrate the effectiveness of these improvements through rigorous experiments. Additionally, we conduct extensive experimental and theoretical analyses to address two pivotal research questions in large-scale TTS: (1) identifying the characteristics of an effective speech tokenizer, and (2) understanding why ASR transcriptions are suitable for training TTS systems. Our findings confirm that SimpleSpeech 2 is not only simple and efficient but also stable, capable of synthesizing speech that is diverse, natural, and controllable.
\bibliographystyle{ieeetr}
\balance
\bibliography{refs.bib}
\end{document}